\documentclass[pre, twocolumn]{revtex4}
\usepackage{amsmath}
\usepackage{amsfonts}
\usepackage{lipsum}
\usepackage{braket}
\usepackage{bm}
\usepackage[dvipdfmx]{graphicx}
\usepackage{xcolor}

\newcommand{\ignore}[1]{}
\newcommand{\beq}{\begin{equation}}
\newcommand{\eeq}{\end{equation}}

\begin{document}
\title{Optical Bistability in a Low Photon-Density Regime}
\author{Tatsuhiko Shirai$^1$}
\email{shirai@exa.phys.s.u-tokyo.ac.jp}
\author{Synge Todo$^{2,1}$}
\email{wistaria@phys.s.u-tokyo.ac.jp}
\author{Hans de Raedt$^3$}
\email{h.a.de.raedt@rug.nl}
\author{Seiji Miyashita$^2$}
\email{miya@spin.phys.s.u-tokyo.ac.jp}
\affiliation{%
$^1$The Institute for Solid State Physics, University of Tokyo, 5-1-5 Kashiwanoha, Kashiwa, Chiba 277-8581, Japan\\
}
\affiliation{%
$^2$Department of Physics, Graduate School of Science,
The University of Tokyo, 7-3-1 Hongo, Bunkyo-Ku, Tokyo 113-0033, Japan\\
}
\affiliation{%
$^3$Department of Applied Physics, Zernike Institute for Advanced Materials,
University of Groningen, Nijenborgh 4, NL-9747AG Groningen, The Netherlands\\
}

\begin{abstract}
We give a microscopic description of the optical bistability, where the transmission coefficient has two different values as a function of input light intensity, and the system exhibits a discontinuous jump with a hysteresis loop.
We developed an efficient numerical algorithm to treat the quantum master equation for hybridized systems of many photons and a large number of two-level atoms.
By using this method, we characterize the bistability from the viewpoint of eigenmodes and eigenvalues of the time evolution operator of the quantum master equation.
We investigate the optical bistability within the low photon-density regime, where the hybridization of photon and atom degrees of freedom occurs and the resonance spectrum has a double peak structure.
We compared it with the standard optical bistability between the low photon-density regime and the high photon-density regime, where the photons can be treated as a classical  electromagnetic field and the resonance spectrum has a single peak structure.
We discuss the steady-state properties of the optical bistability: dependencies of the photon number density on the intensity and the double peak structure of the photon number distribution inside the bistable region.
As for the dynamical properties, we find that the relaxation timescale shows an exponential growth with the system size, and reveal how the hysteresis loop of the optical bistability depends on the size of the system and the sweeping rate of the driving amplitude. 
Finally, by investigating the effects of detuning frequency of the input field, we clarify the characteristic properties of the present optical bistability within the low photon-density regime, which are qualitatively different from the standard optical bistable phenomena.
\end{abstract}

\maketitle
\section{Introduction}\label{intro}
The interplay of atom degrees of freedom and photon degrees of freedom in a microcavity attracts much interest for decades.
The cavity system can be modeled by the Rabi model or the Dicke model~\cite{dicke1954coherence},
consisting of one or a number of two-level atoms coupled to a boson mode.
And often, by adopting the rotational wave approximation (RWA), the Jaynes-Cummings model~\cite{jaynes1963comparison,shore1993jaynes} or the Tavis-Cummings model~\cite{tavis1968exact} has been studied well in order to elucidate the interplay between photons and atoms.

The optical response of atomic systems is qualitatively different depending on whether the photon density is low or high compared with that of atoms.
It has been pointed out that
the crossover between the two regimes occurs when the number of atoms, $N$, is about the same as the number of cavity photons, $n$~\cite{miyashita2012photon}.
When the number of photons inside the cavity is small, i.e., $n < N$, a hybridization of photon and atom degree of freedoms appears in the emission spectrum as a double peak structure.
The Agarwal vacuum-field Rabi splitting is a typical example~\cite{agarwal1984vacuum}.
Such double peak structures due to the hybridization have been found in various experiments~\cite{thompson1992observation,weisbuch1992observation,wallraff2004strong,chiorescu2010magnetic,kubo2010strong} and drawn a lot of attention as a possible memory mechanism to store the photon quantum state in a state of material (quantum RAM)~\cite{blencowe2010quantum}.
We call this region ``low photon-density regime''.
On the other hand, when the number of photons is large, i.e., $n > N$, photons behave as a classical electromagnetic field.
In this case, the population dynamics of the atomic system is given by the standard Rabi oscillation, which gives a single peak in the ESR spectrum of photon absorption.
We call this region ``high photon-density regime''.

Due to the interplay between photons and atoms,
the system exhibits various dynamical phase transitions depending on the strength of the driving field~\cite{shirai2013novel},
optical bistability being one of the well-known examples~\cite{lugiato1984ii}.
The optical bistability manifests itself as a discontinuous transition between a state with high transmission and a state with low transmission.
Bistable nature of the transmission was first observed in experiments in continuum materials such as atomic gases and semiconductor solid state systems~\cite{gibbs1976differential,felber1976theory}.
In these systems, classical electromagnetic theories, such as the Maxwell-Bloch equation~\cite{bonifacio1978optical}, describe the phenomenon well.

The finite-size, i.e., finite-$N$, effects of the optical bistability have also been investigated in experiments.
Progress in cavity quantum electrodynamics (QED) and circuit QED experiments has realized systems with $N<100$.
There, discontinuous behavior of the transmission and size dependence of the hysteresis loop were investigated~\cite{rempe1991optical}.
Besides the bistable nature of the transmission on the driving amplitude, it was found that the transmission spectrum for detuned driving frequency changes from a double peak structure to a single peak.
Furthermore, metastable structures of the spectrum were also studied~\cite{gripp1996anharmonicity}.
Recently, systems with a few atoms have been realized, and there the system is controlled with single-atom resolution.
Optically bistable states have been observed even in such small systems~\cite{kerckhoff2011remnants}.

Extensive efforts have also been devoted to the theoretical side.
The microscopic description of optically bistable phenomena was proposed in Ref.~\cite{bonifacio1978optical}.
The dynamics is described by the quantum master equation (QME), in which the coherent atom-photon coupling and dissipative effects are taken into account.
The bistable features have been explained by a mean-field (MF) treatment.
Indeed, the long-range nature of the interaction between atoms via photons justifies the MF treatment in the limit of $N \to \infty$~\cite{mori2013exactness}.
Finite-size effects have been taken into account by mapping the QME onto a classical equation, such as the Fokker-Planck equation~\cite{gronchi1978fokker} or the Langevin equation~\cite{gardiner2004quantum}.
However, in these mappings, the expansion about the inverse system size is truncated up to the second order or the quantum noise is replaced by a white Gaussian noise.
Therefore, these approximations are valid for the timescale of the order of $N^2$.
In such treatments, the transition process between the optically bistable states is not fully taken into account since its timescale is expected to be $\exp[O(N)]$.
Thus, in order to capture the system size dependence of the optical bistability correctly, a fully quantum description without using such approximations is necessary.

There are a few numerical studies on the optical bistability using the QME for relatively small systems.
It was found that even in the case of a single atom, a double peak structure of photon number distribution was observed~\cite{savage1988single}, the position of two peaks being associated to the bistability.
In Ref.~\cite{dombi2013optical}, the size dependence has been investigated up to $N=8$ with the upper limit of the photon number, $n_{\rm  max}=200$.
In these works, one of the peaks is located at $n<N$ and the other is at $n>N$.
This indicates that the transition occurs between the low photon-density regime and the high photon-density regime, which is the case of the standard optical bistability. 

The optical bistability is expected from the MF theory even in a system with lower photon density, in which the high photon-density state of the optical bistability is still in the low photon-density regime, $n < N$.
In this case, for small systems the signature of the bistability is smeared out.
Indeed, the double peak structure of the photon number distribution was not reported so far~\cite{sarkar1987optical,rice1988single}, though this case would be also important for the manipulation of the photon state in the ultra-low radiation regime.
The larger number of atoms is necessary to observe the optical bistability in this regime.

In this paper we focus on this low photon-density regime, and have developed a computational scheme to solve the QME that treats systems with large number of atoms.
The scheme of numerical calculation consists of the parallelization in photon space by making use of the fact that the time-evolution operator of the QME, ${\cal L}$, is a sparse matrix.
For the Hilbert space representing the atom, we use the permutation symmetry of $\cal{L}$, by which we can reduce the number of dimensions drastically from $2^{2N}$ to $O(N^3)$~\cite{sarkar1987optical,lee2012collective,gegg2016efficient}.
In this scheme, we could in principle study up to the system with $N=100$ and $n_{\rm max}=800$ by using the state-of-the-art supercomputers (see Appendix~\ref{APP2}).
The photon number $n$ should be infinite in principle, but we found that the system is well described if we set $n_{\rm max}$ to be larger than a few times of $N$, as we will see later.

By using the method mentioned above, we first study the steady-state properties of the bistability in the low photon-density regime.
We obtain the photon number density as a function of the amplitude of the driving field, and find that it converges to the MF result as $N$ is increased.
We also investigate the steady-state density matrix, and find a double peak structure in the photon number distribution. 
To clarify the bistable nature, we also analyze the steady-state density matrix by the eigenmode decomposition.

In addition, we study the dynamical aspects of the optical bistability. 
The relaxation time is evaluated from the eigenvalue spectrum of ${\cal L}$, and it is found that in the bistable region, the relaxation time exhibits an exponential growth with the system size.
We also study the hysteresis associated with the optical bistability and obtain its dependence on $N$ and on the sweeping rate of the driving field amplitude.
We show that the relaxation time and the hysteresis loop show the same size dependencies.

We also point out a characteristic of the optical bistability in the low photon-density regime by studying the transmission spectrum for the detuned driving frequency.
In contrast to the standard case, it is found that the spectrum still has a double peak even in the high photon-density state of the optical bistability.
And the size dependence of the spectrum is also studied.

The rest of this paper is organized as follows:
Sec.~\ref{model} gives the microscopic model to describe the optical bistability.
In Sec.~\ref{Sec3:MF}, the scaling of quantities and the MF method are explained.
In Sec.~\ref{simulation}, we explain our numerical method, which allows us to investigate systems with a large number of atoms.
In Sec.~\ref{static}, we study the size dependencies of static properties of the optical bistability.
In Sec.~\ref{dynamic}, we study the size dependencies of dynamic properties, and show the relation with the hysteresis loop.
In Sec.~\ref{Sec:detuning}, we further investigate an effect of the detuned driving frequency, which is a characteristic of the optical bistability in the low photon-density regime.
Finally, the paper is summarized in Sec.~\ref{summary}.

\section{Microscopic model}~\label{model}
\begin{figure}[t]
\includegraphics[clip,width=0.5\textwidth]{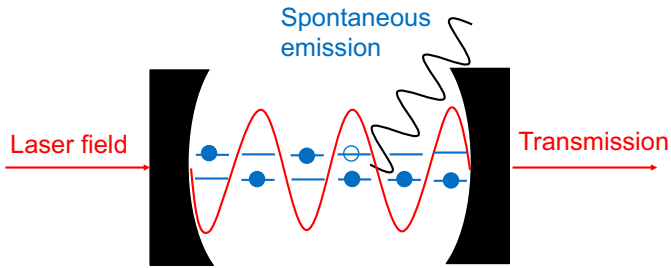}
\caption{
Schematic picture of a cavity system.
The ensemble of two-level atoms interacts with a single quantized mode of the cavity field, which is driven by a laser field.
Transmission of light from the cavity and spontaneous emission of  atoms are taken into account as dissipation.
}
\label{schematic}
\end{figure}

The optical bistability appears in a cavity system with a coherent driving and dissipation  (see Fig~\ref{schematic}).
In order to describe the quantum dynamics of the system, we consider the following QME,
\begin{equation}
{d\rho(t)\over dt}=-i\left[ H(t), \rho (t)\right] + D[\rho(t)]. \label{QME}
\end{equation}
The first term represents the time evolution of the density matrix, $\rho(t)$, under the system Hamiltonian, $H(t)$, and the second term describes dissipative effects.
In this paper we omit $\hbar$ for simplicity.

The Hamiltonian for the cavity system is divided into a static part, $H_0$, and a driving part, $H_{\rm ex}(t)$,
\begin{equation}
H(t)=H_0+H_{\rm ex}(t).\label{cavity}
\end{equation}
The static part represents the cavity system consisting of photons and $N$ atoms with discrete energy levels and is described by the Dicke model~\cite{dicke1954coherence}:
\begin{equation}
H_0 = \omega_{\rm ph} a^{\dagger} a + \sum_{i=1}^N \omega_{\rm a} S_i^z
+i\tilde{g}(a^{\dagger}-a)\sum_{i=1}^N (S_i^++S_i^-),
\label{eqn:dicke}
\end{equation}
where $\omega_{\rm ph}$ is the frequency of the cavity mode.
Here, we confine ourselves to the case of two energy levels per atom and represent the atomic state by a spin-1/2 operator, $\bm{S}_i=\{S_i^x, S_i^y, S_i^z\}$.
The raising and lowering operators are defined by $S_i^{\pm}=S_i^x\pm i S_i^y$.
Hereafter, we call the atom with the discrete energy levels `spin'.
The energy gap between the two states is denoted by $\omega_{\rm a}$.
The interaction between photons and spins is given by the third term in Eq.~(\ref{eqn:dicke}).
The coefficient $\tilde{g}$ is the strength of the interaction.

For the driving part, we adopt the following form:
\begin{equation}
H_{\rm ex}(t)=i\tilde{\xi} \left(a^{\dagger}e^{-i\Omega t} - a e^{i\Omega t} \right),
\label{ham-xi}
\end{equation} 
where $\tilde{\xi}$ and  $\Omega$ are the amplitude and the frequency of the driving field, respectively.
In the present work, we suppose that the energy of a cavity photon and a two-level atom to be the same:
\begin{equation}
\omega_{\rm a}=\omega_{\rm ph}\equiv \omega, \label{unit}
\end{equation}
and set $\omega$ as the unit of the energy.
However, the driving frequency may be detuned by $\Delta \omega$:
\begin{equation}
\Omega=\omega-\Delta\omega. \label{detune_eq}
\end{equation}
We mainly consider the resonant case, $\Delta \omega =0$, except in Sec.~\ref{Sec:detuning}.

For the dissipative term in Eq.~(\ref{QME}), we adopt a standard Lindblad form:
\begin{align}
D[\rho(t)]=&\kappa\left[ 2a\rho a^{\dagger} - (a^{\dagger}a\rho+\rho a^{\dagger}a)
\right]\nonumber\\
&+\gamma \sum_{i=1}^N\left[2 S_i^{-}\rho S_i^{+} - (S_i^+ S_i^{-}\rho+\rho S_i^+S_i^{-}) \right],
\end{align}
where the first term is for the photon transmission from the cavity, and the second term is for the spontaneous emission of each atom.
We consider independent baths for each atom and photons, and therefore the total angular momentum, $\sum_{i=1}^N \bm{S}_i$, is not conserved.

The Lindblad terms are derived by combining the Born-Markov approximation and the secular approximation,
which are justified as long as $\tilde{g}$ and $\tilde{\xi}$ are comparable with the dissipative strength, $\kappa$ and $\gamma$, and much smaller than the resonance frequency, $\omega$.
The optical bistability has been observed in the regime where the above approximations are applicable, 
and thus we expect that the Lindblad form suffices to describe the qualitative nature of the bistable phenomena, though
for general cases where $\tilde{g}$ and/or $\tilde{\xi}$ are comparable with $\omega$, the effects of the atom-photon coupling and the driving field should be incorporated in the dissipation in order to describe the steady state even qualitatively~\cite{shirai2013novel}.

In the present model, we assume uniform couplings between photons and spins, the same dissipative effect for each spin, and no direct interaction among spins.
This property is useful to reduce the size of the density matrix $\rho (t)$ as we will see in Sec.~\ref{simulation}~\cite{sarkar1987optical,lee2012collective,gegg2016efficient}.

In order to simplify the equation further, we use the RWA.
Namely, we work in the rotating frame, in which the density matrix $\rho_{\rm R}(t)$ is given by
\begin{equation}
\rho_{\rm R} (t)=U(t) \rho(t) U^{\dagger}(t), \quad U(t)=e^{-i\omega t (a^\dagger a +\sum_{i=1}^N S_i^z)}.
\end{equation}
The Hamiltonian in the rotating frame reads
\begin{align}
  H_{\rm R}(t)=&U^{\dagger}(t) \left( H(t) -i \frac{\partial}{\partial t} \right) U(t), \nonumber\\
=&i\tilde{g}\sum_{i=1}^N (a^{\dagger} S_i^- -a S_i^+) +i \tilde{\xi} (a^{\dagger}-a) \nonumber\\
&+i\tilde{g}\sum_{i=1}^N (a^{\dagger}S_i^+ e^{2i\omega t} -a S_i^- e^{-2i\omega t}).
\end{align}
In the RWA, we drop the last term in $H_{\rm R} (t)$.
The RWA is not valid in the ultra-strong coupling regime, $\tilde{g} \sim \omega$, and/or under strong driving field, $\tilde{\xi} \sim \omega$~\cite{shirai2013novel},
but  it gives a qualitatively correct behavior in the parameter region for the optical bistability.
Then, the Hamiltonian in the rotating frame becomes time independent,
\begin{equation}
H_{\rm R}=i\tilde{g}\sum_{i=1}^N (a^{\dagger} S_i^- -a S_i^+ ) + i\tilde{\xi} (a^{\dagger}-a),\end{equation}
and the QME in the rotating frame reads
\begin{equation}
{d\rho_{\rm R}(t)\over dt}=-i\left[ H_{\rm R}, \rho_{\rm R} (t)\right] + D[\rho_{\rm R}(t)]\equiv {\cal L} \rho_{\rm R}(t).
\label{QME_R}
\end{equation}
It is noted that the form of the dissipative term does not change under the RWA.
In the following, we use $\rho$ instead of $\rho_{\rm R}$ for simplicity.
Equation (\ref{QME_R}) defines the linear operator ${\cal L}$ and due to the time independence of ${\cal L}$ the steady state is defined by
\begin{equation}
{\cal L} \rho_{\rm ss}=0.
\label{steady_state}
\end{equation}

\section{Mean-field analysis}~\label{Sec3:MF}
In the present model, due to the uniform coupling between photons and spins,
the MF approximation becomes exact for $N \to \infty$ with an appropriate scaling of $\tilde{g}$ and $\tilde{\xi}$~\cite{mori2013exactness}.
For the scaling of the coupling constant $\tilde{g}$,
it is noted that $\tilde{g}$ is usually proportional to $1/\sqrt{V}$, where $V$ is the volume of the cavity.
When the atoms distribute uniformly inside the cavity with a fixed number density $\rho=N/V$, $\tilde{g}$ is proportional to $1/\sqrt{N}$.
Thus we set 
\begin{equation}
\tilde{g}\equiv{g\over \sqrt{N}}
\end{equation}
with an $O(1)$ parameter, $g$.
For the scaling of the driving amplitude $\tilde{\xi}$, on the other hand, it should be scaled as $\sqrt{N}$ in the large $N$ limit,
and thus we set
\begin{equation}
  \tilde{\xi}\equiv{\sqrt{N}\,\xi},
  \label{eqn:scale_xi}
\end{equation}
where $\xi$ is independent of $N$, i.e., $O(1)$.
The expectation value of the photon number in
 the steady state is given by
\begin{equation}
n={\rm Tr} (a^\dagger a \rho_{\rm ss}).
 \label{photon_density}
\end{equation}
In this scaling, the photon number density, $n/N$, is proportional to $\xi^2$, which is independent of $N$.

In the MF approximation, the density matrix is assumed to be given in the product form:
\begin{equation}
\rho(t)=\rho_{\rm ph}(t) \otimes \rho_{\rm s}^{\otimes_N}(t),
\end{equation}
where $\rho_{\rm ph}(t)$ and $\rho_{\rm s}(t)$ are the density matrices of the photon and spin, respectively.
Here we assume the density matrix of each spin to be the same for all the spins.
Substituting this product form into Eq.~(\ref{QME_R}), we obtain the closed set of equations:
\begin{equation}
\left\{
\begin{aligned}
&{\partial \alpha\over\partial t}=(gm^- + \xi)-\kappa \alpha, \\
&{\partial m^-\over\partial t}=2g \alpha m^z -\gamma m^-, \\
&{\partial m^z\over\partial t}=-g(\alpha^* m^- + \alpha m^+)-\gamma(2 m^z+1),
\end{aligned}
\right.
\end{equation}
where
\begin{equation}
\alpha (t) \equiv {{\rm Tr} [a\rho (t)]\over \sqrt{N}}
\end{equation}
and
\begin{equation}
m^{\pm}={\rm Tr}[S_i^{\pm} \rho(t)], \quad m^{z}={\rm Tr}[S_i^{z} \rho(t)].
\end{equation}
This is the MF equation for the optical bistability originally given in Ref.~\cite{bonifacio1978optical}.

\begin{figure}[t]
\includegraphics[clip,width=0.5\textwidth]{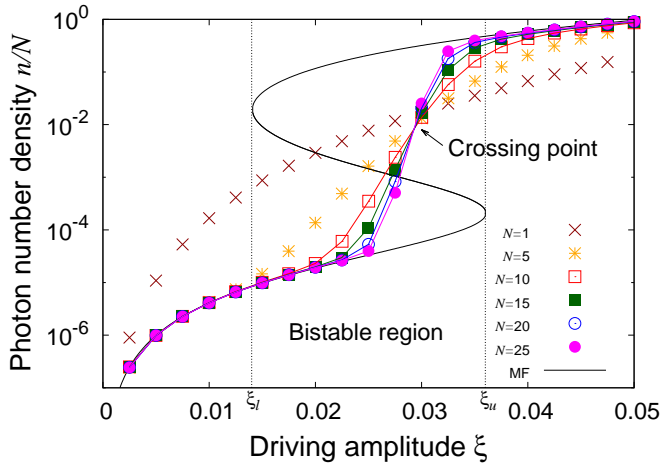}
\caption{
Dependence of photon number density $n/N$ on the driving amplitude $\xi$ for various values of $N$ for $g=0.1$, $\kappa=0.05$, and $\gamma=0.002$ $(C=50)$.
The solid line denotes the MF solution [Eq.~(\ref{MF})], which exhibits multiple stationary solutions in the bistable regime, $\xi_{\rm l}<\xi<\xi_{\rm u}$.
The symbols show values obtained by the numerical method (explained in Sec.~\ref{simulation}) for $N=1,5,10,\cdots,25$.
}
\label{OB}
\end{figure}

The MF solution for the steady state is obtained by setting the r.h.s of these equations to be zero, from which the following relation between $\xi$ and $\alpha$ is obtained:
\begin{equation}
\xi=\left({\kappa+{\gamma g^2\over 2g^2\alpha^2+\gamma^2}}\right)\alpha 
 \label{MF}.
\end{equation}
With the scaled parameters,
\beq
\alpha_{\rm s}={\sqrt{2}g\over \gamma}\alpha, \quad \xi_{\rm s}={\sqrt{2}g\over \kappa\gamma}\xi, \label{scale}
\eeq
the relation reads
\beq
\xi_{\rm s}=\left(1+{2C\over \alpha_{\rm s}^2+1}\right)\alpha_{\rm s}, \label{OB_scale}
\eeq
where $C$ is the cavity cooperativity parameter:
\beq
C={g^2\over 2\kappa\gamma}.\label{cooperativity}
\eeq
It should be noted that the cases with the same $C$ give the same dependence between $\xi_{\rm s}$ and $\alpha_{\rm s}$, and the bistable states appear when $C>4$.
But the photon number density $n/N$ itself depends on $\kappa$ and $\gamma$ as 
\beq
{n \over N}=\alpha^2 = {\gamma \over 4 \kappa C}\alpha_{\rm s}^2. \label{density_scale}
\eeq
The solution is depicted by the solid line in Fig.~\ref{OB} for the case $g=0.1$, $\kappa=0.05$, and $\gamma=0.002$ $(C=50)$, in which the bistable region appears for
\begin{equation}
\xi_{\rm l} \equiv 1.40705\times10^{-2} < \xi < \xi_{\rm u} \equiv 3.60697 \times 10^{-2}.\label{bistable}
\end{equation}
It is noted that with the present set of parameters the high photon-density state is still in the low photon-density regime, i.e., $n/N < 1$.
In the figure, we also plot the data obtained by the numerical method, to be discussed later.

\section{Numerical methods}~\label{simulation}
In this section we explain our numerical methods to study the properties of the system given by the QME [Eq.~(\ref{QME_R})].
The QME is a linear equation of the density matrix $\rho(t)$
and therefore all the properties are obtained by solving the eigenvalue problem of the linear operator ${\cal L}$.
The steady state corresponds to the eigenmode $\rho_{1}$ with zero eigenvalue $(\lambda_1\equiv 0)$ of ${\cal L}$. 
We denote the mode $\rho_{1}$ by $\rho_{\rm ss}$ because it obeys the equation for the steady-state density matrix, i.e., Eq.~(\ref{steady_state}).

In the present numerical calculation, we rewrite $\rho$ as a vector $\vec{\rho}$.
Since $\rho$ is a matrix of $[(n_{\rm max}+1) \Omega_{\rm spin}]\times [(n_{\rm max}+1) \Omega_{\rm spin}]$, where $n_{\rm max}$ is a cutoff for the photon number, $\vec{\rho}$ is a $[(n_{\rm max}+1) \Omega_{\rm spin}]^2$-dimensional vector.
Here $\Omega_{\rm spin}^2$ is $2^{2N}$ for general cases, but in the present case it is reduced to be $O(N^3)$ by using the symmetry as we will see in Eq.~(\ref{Ndim}).
In the vector representation, the time evolution operator ${\cal L}$ is expressed as a $[(n_{\rm max}+1) \Omega_{\rm spin}]^2 \times [(n_{\rm max}+1) \Omega_{\rm spin}]^2$ matrix $L$, then, the QME is expressed as
\begin{equation}
  {d\over dt}\vec{\rho}=L\vec{\rho}.
  \label{matrix_rep}
\end{equation}
The number of nonzero matrix elements of $L$ is the order of $[(n_{\rm max}+1) \Omega_{\rm spin}]^2$. 
In our simulation, we prepare a list of non-zero elements to perform the product of the sparse matrix $L$ and the vector $\vec{\rho}$ efficiently.
Moreover, the amount of required memory for $L$ can also be reduced to the order of $[(n_{\rm max}+1) \Omega_{\rm spin}]^2$.

\begin{figure}[t]
\includegraphics[clip,width=0.5\textwidth]{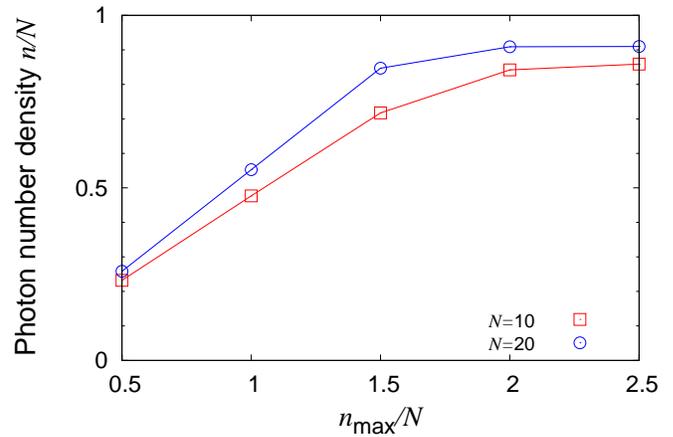}
\caption{
Photon number density $n/N$ in a restricted photon space, which is specified by the photon number cutoff $n_{\rm max}$ for $N=10$ (red squares) and $N=20$ (blue circles).
The values of parameters are given by $(g,\xi,\kappa,\gamma)=(0.1,0.05,0.05,0.002)$.
}
\label{nmax}
\end{figure}

The photon space is labeled by the photon number $n$, i.e., 
\beq
a^{\dagger} a \ket{n} =n \ket{n}.
\eeq
In principle, the photon number $n$ runs from $0$ to $\infty$, but in numerical calculations, we find that if we take the photon number cutoff $n_{\rm max}$ sufficiently large, the numerical data converges.
It is noted that $n_{\rm max}$ becomes larger as $N$ increases.
We found that it is necessary to set $n_{\rm max}$ larger than a few times of $N$.
In Fig.~\ref{nmax}, we show the dependence of the photon number density $n/N$ [see Eq.~(\ref{photon_density})] as a function of $n_{\rm max}$ for $N=10$ and $20$.

It should be noted that the present model has a permutation symmetry of spins: all the spins interact with each other via a common photon field, and $g$ and $\gamma$ are the same for all the spins.
In such a case, we can reduce the dimension of spin space by making use of the symmetry~\cite{sarkar1987optical,lee2012collective,gegg2016efficient}, i.e.,
\beq
\begin{split}
  \Omega_{\rm spin}^2 & =2^{2N} \\ \rightarrow & N_{\rm dim}\equiv{(N+1)(N+2)(N+3)\over 6} = O(N^3).
\end{split}
\label{Ndim}
\eeq 

\begin{figure}[t]
\includegraphics[clip,width=0.5\textwidth]{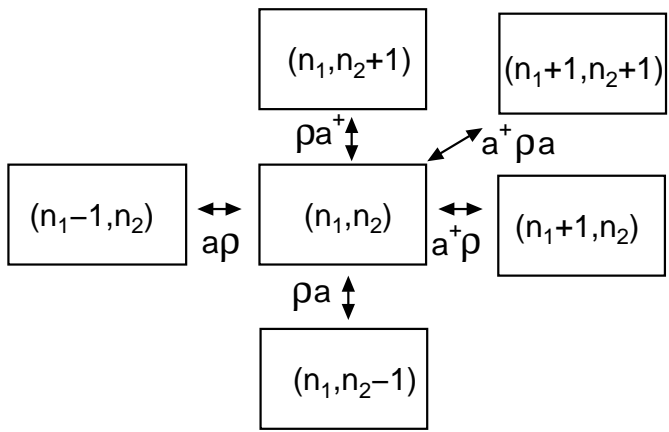}
\caption{Data exchange among neighboring cores necessary for the multiplication of $L$ on $\vec{\rho}$ in the parallelization method.}
\label{cores}
\end{figure}

Thus, the total dimension of $\vec{\rho}$ becomes $(n_{\rm max}+1)^2 N_{\rm dim}$, which is still too large to fit in a single core of a typical computer.
In the present work, we adopt the distributed-memory parallelization on a supercomputer, which enables us to reduce the memory requirement on each core significantly.
Specifically, we label the elements of $\vec{\rho}$ by two photon numbers $n_1$ and $n_2$, which corresponds to $\bra{n_1} \rho \ket{n_2}$, and assign them to different cores.
Each core stores $N_{\rm dim}$ elements for spin states.
In the present system, $L$ is sparse in the photon space.
Indeed, the multiplication of $L$ and $L^{\dagger}$ to $\vec{\rho}$ only requires exchange of data between the neighboring cores (see Fig.~\ref{cores}), because the operations change the photon number only by $\pm 1$, e.g., for the calculation of $(n_1, n_2)$-elements of $a \rho$, only the $(n_1+1, n_2)$-elements of $\rho$ are necessary.
In this way, we can achieve good efficiency by the present parallelization scheme (see Appendix~\ref{APP2}).

\subsection{Steady state} 
We obtain $\rho_{\rm ss}$ as the eigenmode with zero eigenvalue of $L$ 
in the space with a finite cutoff of the photon number $n_{\rm max}$.
We solve
\begin{equation}
L\vec{\rho}_{\rm ss}=0
\end{equation}
by the biconjugate gradient (Bi-CG) method~\cite{fletcher1976conjugate}
\footnote{We may obtain the steady state by the Lanczos method for $L^\dagger L$. But the convergence takes much longer time since the gap of eigenvalues between the steady state and the subdominant state becomes much smaller than that for the $L$.}.
It is noted that this equation is homogeneous.
However, we can obtain the steady-state solution with this method because the steady state $\rho_{\rm ss}$ satisfies
\begin{equation}
{\rm Tr}\rho_{\rm ss}=1,
\end{equation}
and the trace of the density matrix is preserved through the iteration process of the Bi-CG method.

\subsection{Relaxation process}
The dynamic properties are related to the subdominant eigenmodes $\{\rho_{i}\}$ with nonzero eigenvalues $\{ \lambda_i \}_{i=2,3.\cdots}$, which satisfy 
${\rm Re}\, \lambda_i < 0$ for $i \geq 2$.
We order the eigenmodes according to
\begin{equation}
0 > {\rm Re}\, \lambda_2 \ge {\rm Re}\, \lambda_3 \ge \cdots.
\end{equation}
In general, the dynamics of the density matrix is given by
\begin{equation}
\rho(t)=\rho_{\rm ss}+\sum_{i=2,3.\cdots}c_ie^{\lambda_i t}\rho_{i}, \label{rho_decompose}
\end{equation}
where the coefficients $\{c_i\}$ are determined by the initial state.
The contribution of each eigenmode with $i \geq 2$ decays as $e^{({\rm Re}\,\lambda_i) t}$ in time. 

The slowest relaxation is governed by the mode with $i=2$, and therefore we define the relaxation time $\tau$ by
\begin{equation}
\tau= -\left({\rm Re}\,\lambda_2 \right)^{-1}. \label{relaxation}
\end{equation}
The value of $\lambda_2$ is estimated by the inverse power method.
In this method,
we first set an initial density matrix $x_0$, and subtract from it the component proportional to the steady state,
\begin{equation}
x_1=x_0-({\rm Tr} x_0) \rho_{\rm ss}.
\end{equation}
Here it is noted that $x_1$ is a traceless matrix, and thus $x_1$ is expanded by $\{\rho_i\}$ with $i \ge 2$~\footnote{The trace-preservation property of $\rho(t)$ in Eq.~(\ref{rho_decompose}) indicates that the eigenmodes $\{\rho_i\}$ of ${\cal L}$ are traceless operator except for $\rho_1\equiv \rho_{\rm ss}$.}.
We then repeatedly solve the following linear equation,
\begin{equation}
{\cal L} x_{k+1} =x_{k} \text{ for } k=1,2,\cdots.
\label{Lxk}
\end{equation}
In order to solve Eq.~(\ref{Lxk}), we again use the Bi-CG method in the vector representation.
The relaxation time is then given by~\footnote{We assume that $\lambda_2$ is real, which is confirmed by exact diagonalization of $L$ for small $N$.}
\begin{equation}
\tau =\lim_{k \to \infty}
\left(\frac{||x_k||_{\rm F}}{||x_{k+1}||_{\rm F}} \right)^{-1},
\end{equation}
where $||\cdot||_{\rm F}$ denotes the Frobenius norm.

\section{Steady-state properties}\label{static}
We performed simulations with the method mentioned above.
We adopt $\omega$ (see Eq.~(\ref{unit})) as a unit of the energy, and we fix the parameters $(g, \kappa, \gamma)=(0.1, 0.05, 0.002)$ as a typical set to study the optical bistability in the low photon-density regime.

\subsection{Photon number density}
We first study the $\xi$-dependence of the photon number density $n/N$ for various system sizes $N$.
As clearly seen in Fig.~\ref{OB}, the steady-state value outside the bistable region quickly converges to the MF value.
On the other hand, deeply inside the bistable region, $n/N$ takes a value between those of the optically bistable states obtained by the MF, and the $\xi$-dependence of $n/N$ becomes shaper and sharper as $N$ is increased.
We also find that the data with different $N$ cross at almost the same point $\xi_{\rm c} \simeq 0.029$.
The steady-state value of $n/N$ for $\xi <\xi_{\rm c}$ approaches the low photon-density state of the MF solution as $N$ is increased, while that for $\xi >\xi_{\rm c}$ does the high photon-density state.
Thus, it is expected that the steady-state value shows a discontinuous jump at the crossing point in the limit of $N \to \infty$.

This behavior is similar to the size dependence of physical quantities of the thermodynamic first-order phase transition, and thus
we call the present observed phenomenon the {\em dynamical first-order phase transition}.
In what follows, we will study this transition from a viewpoint of an effective potential function (a kind of phenomenological free energy) by analyzing the distribution function of quantities which reflect this potential.

\subsection{Photon number distribution in the steady state}
\begin{figure}[t]
\includegraphics[clip,width=0.5\textwidth]{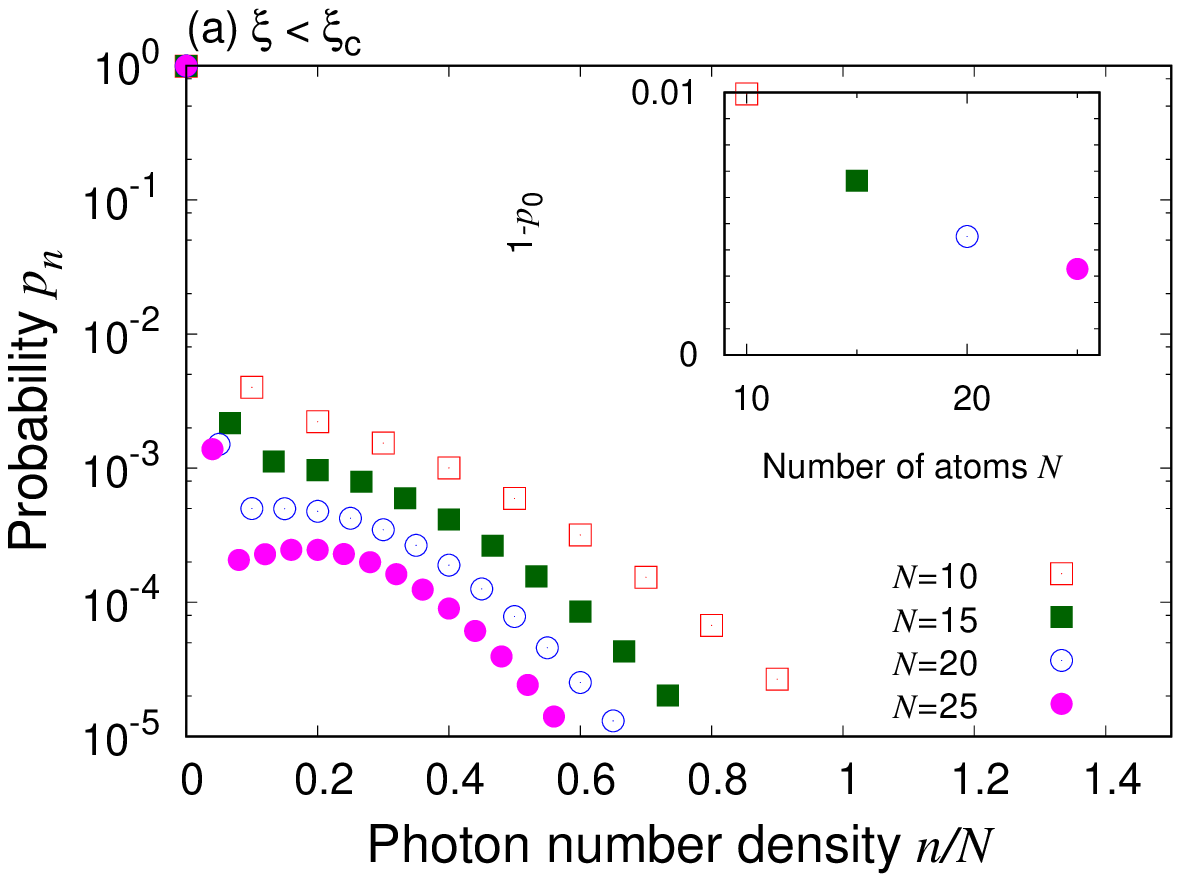}
\includegraphics[clip,width=0.5\textwidth]{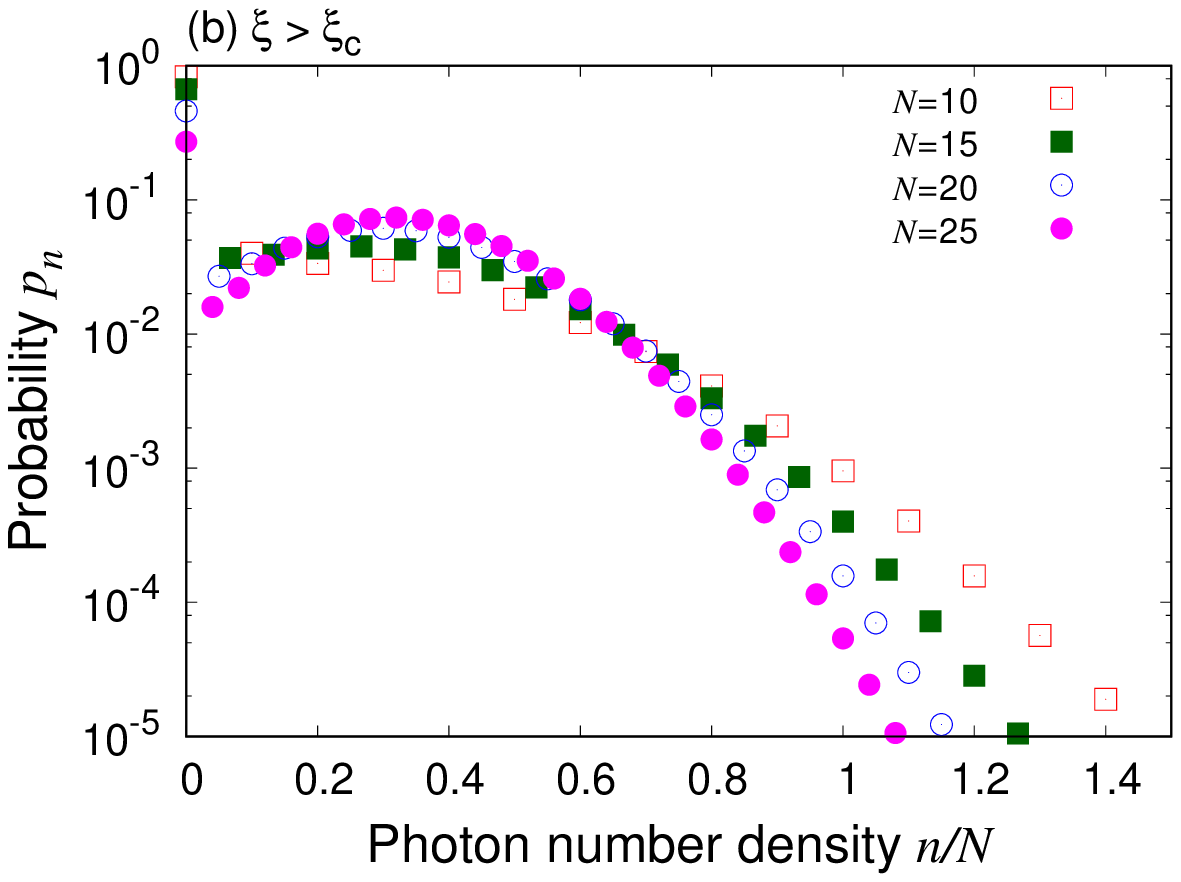}
\caption{
Photon number distribution $p_n={\rm Tr}_{\rm s}\bra{n}\rho_{\rm ss}\ket{n}$ at (a) $\xi =0.0275 < \xi_{\rm c}$ and (b) $\xi = 0.0325 > \xi_{\rm c}$ for $N=10$ (open squares), $15$ (closed squares), $20$ (open circles), and $25$ (closed circles).
The inset of (a) shows the detailed size dependence of $p_0$. Note that $1-p_0$ is plotted instead of $p_0$ in the inset.
}
\label{photon_distribution}
\end{figure}

From Fig.~\ref{OB}, we expect a double peak structure of the photon number distribution in the bistable region.
In this subsection, we study how the photon number distributes in $\rho_{\rm ss}$. 

We define the probability $p_n$ to observe $n$ photons inside the cavity as
\begin{equation}
p_n={\rm Tr}_{\rm s} \bra{n} \rho_{\rm ss} \ket{n}=\bra{n} \rho_{\rm photon} \ket{n},
\end{equation}
where ${\rm Tr}_{\rm s}$ denotes the trace over the spin degrees of the freedom, and the reduced density matrix for photons is defined by 
\begin{equation}
\rho_{\rm photon}\equiv{\rm Tr}_{\rm s} (\rho_{\rm ss}).
\end{equation}
Note that the average number of photons is given by
\begin{equation}
n=\sum_{m=0}^{n_{\rm max}} p_m m.
\end{equation}

We find a double peak structure in the photon number distribution around the crossing point $\xi_{\rm c}$.
We plot $p_n$ as a function of the photon number density $n/N$ at $\xi=0.0275 < \xi_{\rm c}$ [Fig.~\ref{photon_distribution}(a)] and $\xi=0.0325 > \xi_{\rm c}$ [Fig.~\ref{photon_distribution}(b)].
In both cases, one of the peaks is located at $n/N = 0$ and the other is located at a finite photon number density.

However, we find that the size dependencies of the two peaks differ from each other.
In case (a), the peak at $n/N=0$ increases and the other peak at finite $n/N$ decreases with $N$.
In contrast, in case (b), the peak at $n=0$ decreases and the other peak increases with $N$.
This indicates that in the thermodynamic limit, $N \to \infty$, the peak with low photon number density dominates for $\xi < \xi_{\rm c}$, while the peak with high photon number density dominates for $\xi > \xi_{\rm c}$.
The double peak structure has been observed in the standard optical bistability between the low photon-density regime and the high photon-density regime, e.g., Refs.~\cite{lee2012collective,kerckhoff2011remnants}.
In these cases, the photon number density $n/N$ in the high photon-density state is larger than one, and the double peak structure is observed more clearly.
In the present work, we are studying the case where $n/N$ in the high photon-density state is still in the low photon-density regime, i.e., $n/N<1$.
There, the peak at $n=0$ is extremely narrow.
We find that the double peak structure becomes more and more clear as $N$ is increased.

\subsection{Relevant states of the density matrix}
\begin{figure}[t]
\includegraphics[clip,width=0.5\textwidth]{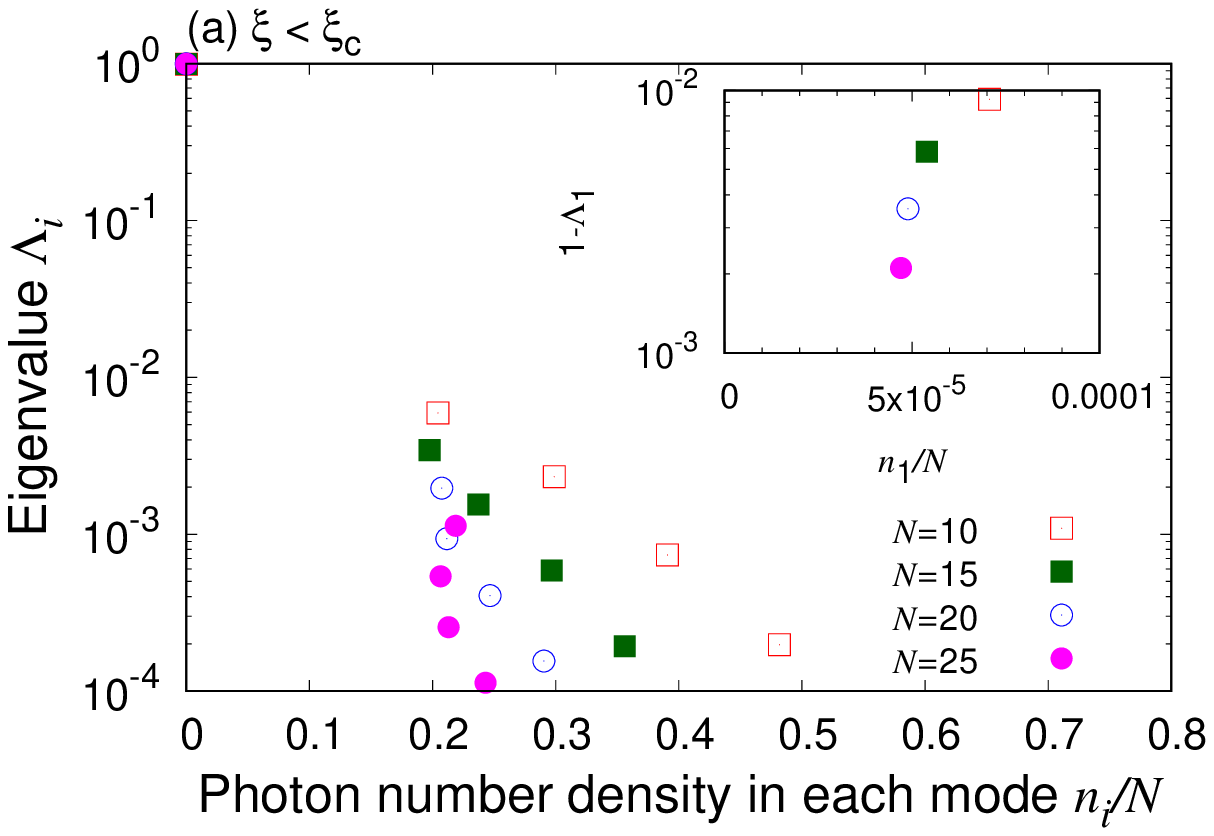}
\includegraphics[clip,width=0.5\textwidth]{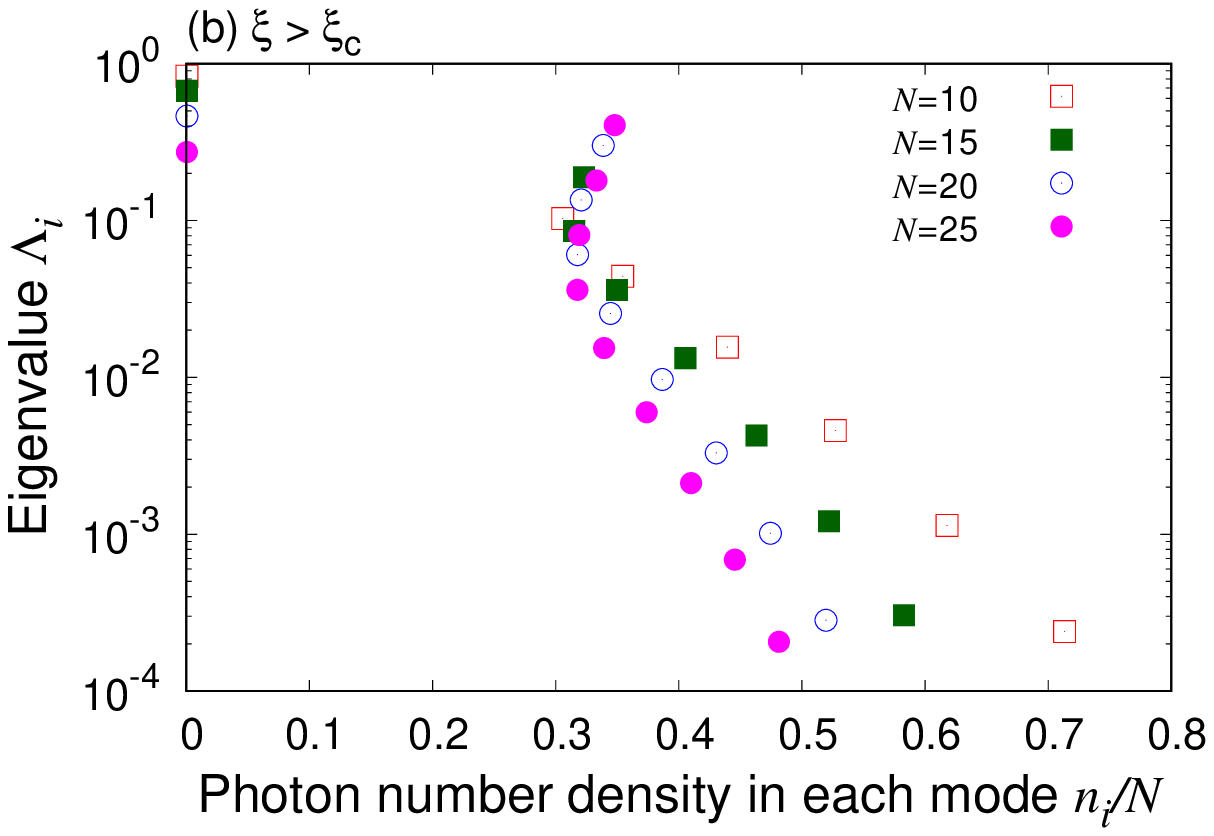}
\caption{
Eigenvalue of each mode, $\Lambda_i$, is plotted as a function of the photon number density $n_i/N$ at (a) $\xi =0.0275 < \xi_{\rm c}$ and (b) $\xi = 0.0325 > \xi_{\rm c}$ for $N=10$ (open squares), $15$ (closed squares), $20$ (open circles), and $25$ (closed circles).
The inset of (a) shows the detailed size dependence of the eigenvalue $\Lambda_1$ located at $n/N \simeq 0$. Note that $1-\Lambda_1$ is plotted instead of $\Lambda_1$ in the inset.
}
\label{distribution}
\end{figure}
In order to grasp the nature of the bistable structure of the steady state of the density matrix,
we perform the eigenmode decomposition:
\begin{equation}
\rho_{\rm photon}=\sum_{i=1}^{n_{\rm max}+1} \Lambda_i \ket{i}\bra{i},
\end{equation}
where $\ket{i}\bra{i}$ denotes the $i$-th mode with eigenvalue $\Lambda_i$.
Here, the index $i$ runs from $1$ to $n_{\rm max}+1$, and we order the eigenmodes in the following manner:
\beq
\Lambda_1\ge \Lambda_2\ge\cdots \ge \Lambda_{n_{\rm max}+1}.
\eeq
The photon number in each mode is given by
\beq
n_i\equiv \bra{i}a^{\dagger} a\ket{i}.
\eeq

In Fig.~\ref{distribution}, we plot $\Lambda_i$ as a function of $n_i/N$ for various system sizes.
In Fig.~\ref{distribution}(a), we find that for $\xi=0.0275<\xi_{\rm c}$, the most dominant mode ($i=1$) with $\Lambda_i\simeq 1$ has almost zero photon ($n_i/N\simeq 5\times 10^{-5}$).
The other modes ($i\ge 2$) have a finite photon number density ($n_i/N\simeq 0.2$), but the eigenvalues of the modes decrease with $N$.
In contrast, in Fig.~\ref{distribution}(b) for $\xi=0.0325>\xi_{\rm c}$,
the eigenvalue of the mode with zero photon decreases with $N$, and the eigenvalues of the modes with finite photon number density increase.
For $N=25$, the eigenvalue at a finite photon number becomes larger than that at zero photon number density, which indicates that in this regime, the most dominant mode is on the high photon density side.
These size dependencies of the double peak structure are consistent with the picture of a first-order phase transition.

\subsection{Effective free energy for dynamical first-order transition}\label{sec:effective}
From the analogy with the static first-order phase transition, we may consider an effective free energy $f(\alpha)$, from which the equation of motion of the order parameter, corresponding to the MF self-consistent equation~(\ref{MF}), is given by
\beq
{df\over d\alpha}=0.
\eeq
Naively, one might expect that from Eq.~(\ref{MF}), we can obtain a candidate of the free energy landscape $\tilde{f}(\alpha)$ by integrating the equation:
\begin{equation}
{d\tilde{f}(\alpha)\over d\alpha}= 
\left({\kappa+{\gamma g^2\over 2g^2\alpha^2+\gamma^2}}\right)\alpha-\xi.\label{free_energy}
\end{equation}
The minima of $\tilde{f}(\alpha)$ reproduce the stable MF solutions $\alpha_1$ and $\alpha_2$ in the bistable region.
However, $\tilde{f}(\alpha)$ does not correctly predict the transition point $\xi_{\rm c}$.
At $\xi_{\rm c}$, the values of $\tilde{f}(\alpha)$ for $\alpha_1$ and $\alpha_2$ are different.
Indeed, the value of $\xi$ where the two minima $\tilde{f}(\alpha_1)$ and $\tilde{f}(\alpha_2)$ are equal with each other is around $\xi \simeq 0.018$, which is different from the crossing point, $\xi_{\rm c} \simeq 0.029$.
In addition, Maxwell's equal area law does not work either.
In this way, the free-energy picture using the MF equation (\ref{MF}) does not work as discussed in Ref.~\cite{lugiato1984ii}.

It should be noted that if we multiply the r.h.s of Eq.~(\ref{free_energy}) by a non-zero smooth function $I(\alpha)$, i.e.,
\beq
{d{f}(\alpha)\over d\alpha}= \left[
\left({\kappa+{\gamma g^2\over 2g^2\alpha^2+\gamma^2}}\right)\alpha-\xi\right]
\times I(\alpha),
\eeq
the position of the minima does not change but the values of the minima do change.
Therefore, there is an ambiguity to find $I(\alpha)$.
We leave the problem to obtain an effective free energy landscape,
\beq
f(\alpha) \propto -{1\over N}\ln \left({\rm Tr}[\delta(\sqrt{a^{\dagger}a}-\alpha)\rho_{\rm ss}] \right),
\eeq
which is a large deviation function of the photon number distribution, for the future study.

\section{Dynamic properties}\label{dynamic}
\subsection{Relaxation time}

\begin{figure}[t]
\includegraphics[clip,width=0.5\textwidth]{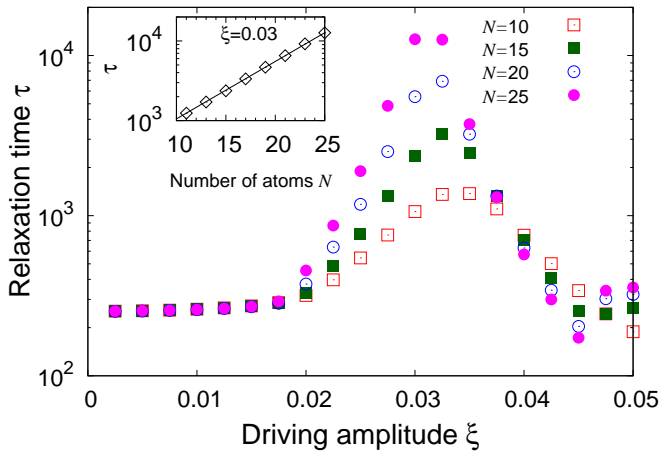}
\caption{
$\xi$-dependence of the relaxation time $\tau$ for $N=10$ (open squares), $15$ (closed squares), $20$ (open circles), and $25$ (closed circles).
Inset: $N$-dependence of $\tau$ at $\xi=0.03$.
}
\label{gap}
\end{figure}
From the double peak structure in the steady-state density matrix,
we expect that the transition probability between the two optically stable states is small.
The smallest transition rate is given by $\lambda_2$, and the relaxation time was defined  in Eq.~(\ref{relaxation}).
If the system has a metastable state, we expect that the relaxation time increases exponentially with $N$ as
\begin{equation}
\tau \sim e^{c N}.
\label{ecN}
\end{equation}
In Fig.~\ref{gap}, we plot the relaxation time $\tau$ as a function of $\xi$.
Around the crossing point $\xi_{\rm c}$, we find that the relaxation time indeed increases exponentially with $N$.
In the inset of the figure, we plot the size dependence of $\tau$ at $\xi =0.03$, which clearly shows the exponential growth with
\begin{equation}
c \simeq 0.166 \quad (\xi \simeq \xi_{\rm c}).
\label{gap_scaling}
\end{equation}
This type of exponential dependence is found to hold around $\xi \simeq \xi_{\rm c}$, but the value of $c$ changes with $\xi$.
This size dependence of the relaxation time is again consistent with the picture of a first-order phase transition. 

\subsection{Hysteresis}
\begin{figure}[t]
\includegraphics[clip,width=0.4\textwidth]{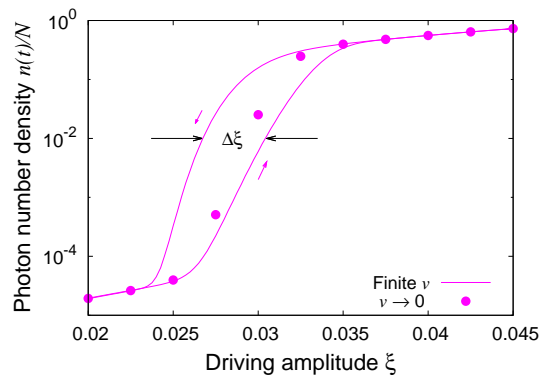}
\caption{ 
An example of the hysteresis loop of photon number density for $N=25$ and $v=2\times 10^{-7}$.
The photon number density in the increasing $\xi$-process and in the deceasing $\xi$-process are depicted by solid lines.
The steady-state values are plotted by filled dots for comparison.
}
\label{hyst25}
\end{figure}

\begin{figure}[t]
\includegraphics[clip,width=0.4\textwidth]{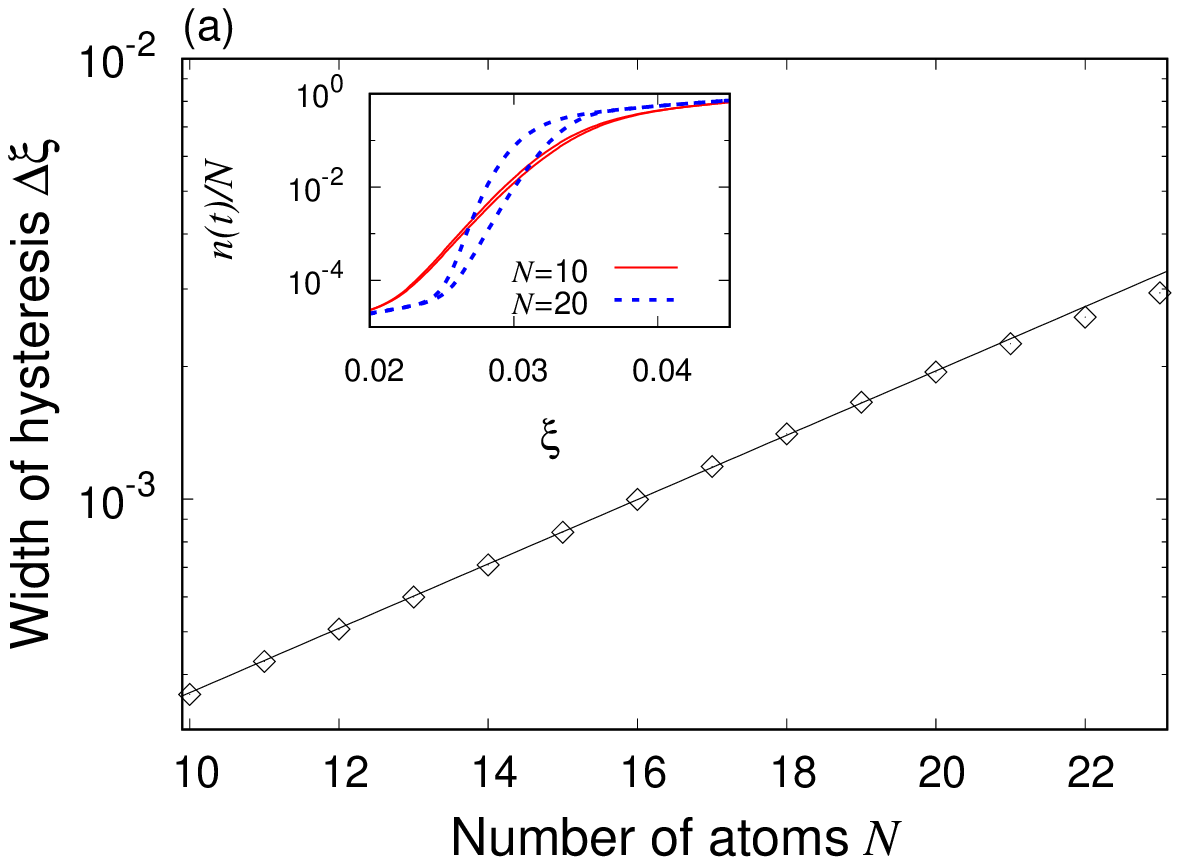}
\includegraphics[clip,width=0.44\textwidth]{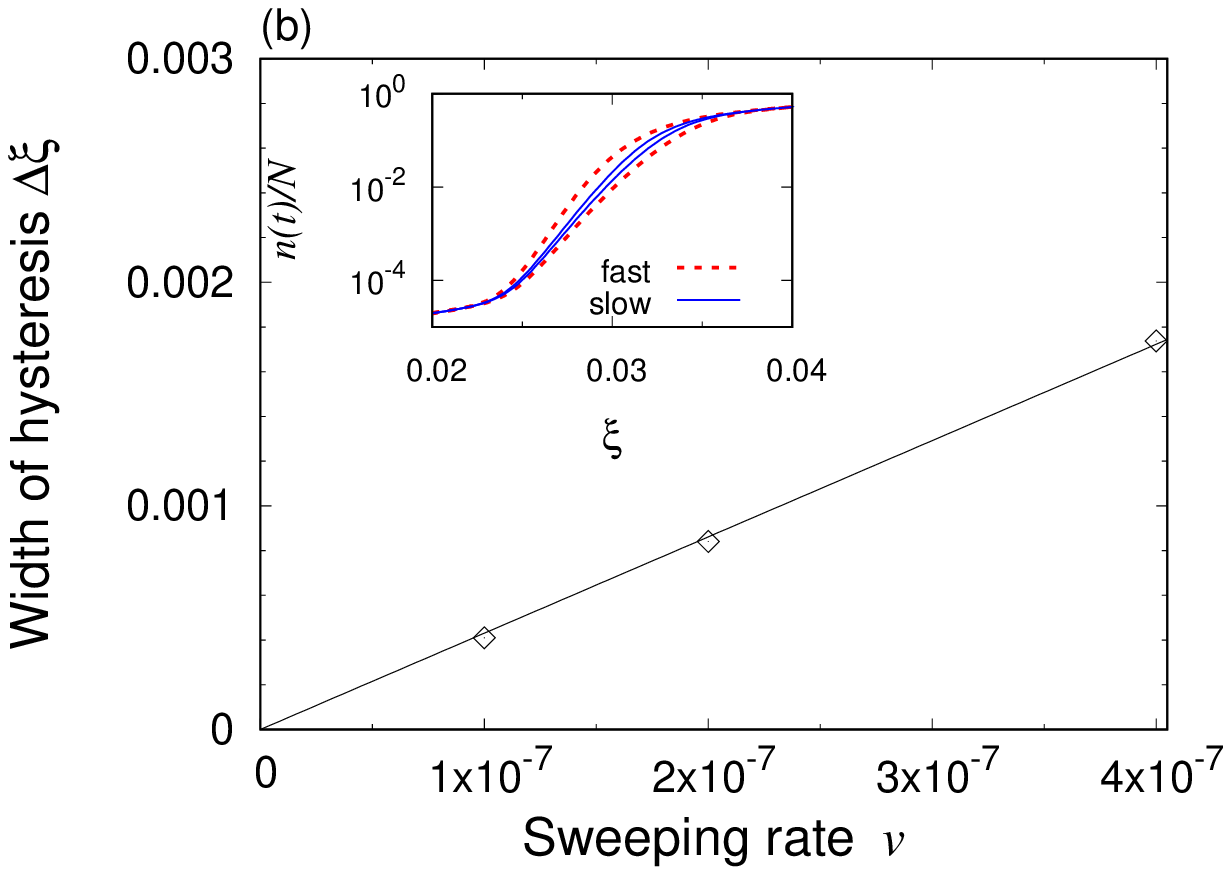}
\caption{
(a) $N$-dependence of the width of hysteresis loop, $\Delta \xi$, for $v=2\times 10^{-7}$.
Inset: the hysteresis loops for $N=10$ (red) and $N=20$ (blue).
(b) $v$-dependence of $\Delta \xi$ for $N=15$.
Inset: the hysteresis loops for $v=4\times 10^{-7}$ (red) and $v=1\times 10^{-7}$ (blue).
}
\label{hysteresis}
\end{figure}

The hysteresis behavior appears when $\xi$ is increased and then decreased at a finite sweeping rate, though we do not see it in the steady state (see Fig.~\ref{OB}).
Here we demonstrate the hysteresis by sweeping $\xi$.
We change $\xi$ from $\xi_{\rm i}=0.02$ to $\xi_{\rm f}=0.045$ at a constant sweeping rate $v$, and then return back to $\xi_{\rm i}$ at the same sweeping rate.
That is, the time dependence of $\xi$ is given by
\begin{equation}
\xi(t)=
\begin{cases}
\xi_{\rm i}+vt & \text{for $0 \leq t \leq T \equiv (\xi_{\rm f} -\xi_{\rm i})/v$} \\
\xi_{\rm f}-v(t-T) & \text{for $T \leq t \leq 2T$}.
\end{cases}
\end{equation}
We set the initial state at $t=0$ to be the steady state for $\xi=\xi_{\rm i}$.
We depict the dynamics of photon number density $n(t)/N$ in this protocol by the solid line in Fig.~\ref{hyst25}.
Here we define the photon number by
\begin{equation}
n(t)={\rm Tr} [a^{\dagger}a \rho(t)],
\end{equation}
where $\rho(t)$ is obtained numerically by solving the QME [Eq.~(\ref{QME_R})] using the parallelized algorithm described above.

The shape of the hysteresis loop depends on the size of the system $N$ and also on the sweeping rate $v$. 
In order to give a quantitative description, we define the width of the hysteresis loop $\Delta \xi$ by the difference of $\xi$ at $n(t)/N=0.01$ in the increasing $\xi$-process and the decreasing $\xi$-process.
In principle, we should define it as the maximum width of the hysteresis loop, but we find that the maximum value is always near  $n(t)/N=0.01$ as shown in the insets of Figs.~\ref{hysteresis} (a) and (b).

The dependencies of $\Delta \xi$ on $N$ and $v$ are depicted in Figs.~\ref{hysteresis} (a) and (b), respectively. We find that $\Delta \xi$ increases with $N$, see Fig.~\ref{hysteresis}(a), and also with $v$, see Fig.~\ref{hysteresis}(b).
We find good linear dependencies in the coordinate $(N, \log \Delta\xi)$ as depicted in Fig.~\ref{hysteresis}(a), and also in $(v,\Delta\xi)$ as depicted in Fig.~\ref{hysteresis}(b).
Thus we conclude that the scaling form
\begin{equation}
\Delta \xi \sim v e^{c' N}
\end{equation}
with $c' \simeq 0.168$ describes the data quite well.
The exponent $c'$ is close to $c \simeq 0.166$ [see Eq.~(\ref{gap_scaling})],
which indicates that the growth of the relaxation time $\tau$ is reflected in the hysteresis loop, and the width of the hysteresis is governed by the slowest relaxation at $\xi_{\rm c}$, where the exponent $c$ becomes maximum.

\section{Detuning effects}\label{Sec:detuning}
\begin{figure}[t]
\includegraphics[clip,width=0.5\textwidth]{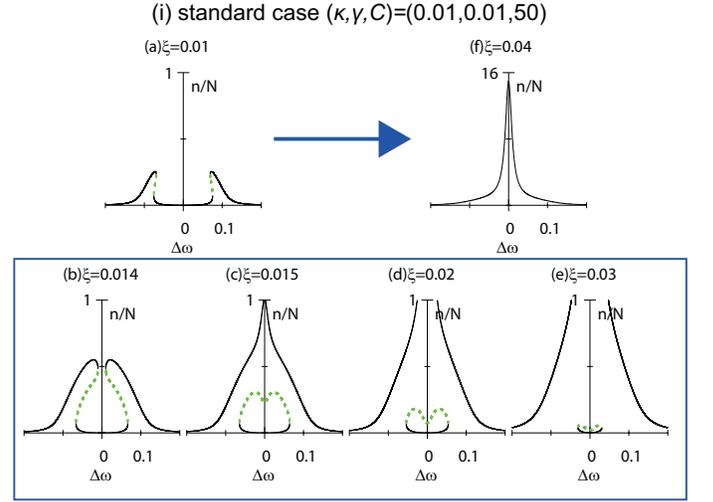}
\caption{
Dependence of the photon number density $n/N$ on the detuning frequency $\Delta \omega$ in the standard case, ($\kappa, \gamma, C) = (0.01, 0.01, 50$), where the single peak appears at $\xi=0.04$.
The transition of the structures from (a) $\xi=0.01$ to (f) $\xi=0.04$ is shown in the box: (b) $\xi =0.014$, (c) $\xi =0.015$, (d) $\xi =0.02$, and (e) $\xi =0.03$.
The stable and unstable MF steady-state solutions are denoted by solid lines (black) and dotted lines (green), respectively.
}
\label{MF-detuning2}
\end{figure}

\begin{figure}[t]
\includegraphics[clip,width=0.5\textwidth]{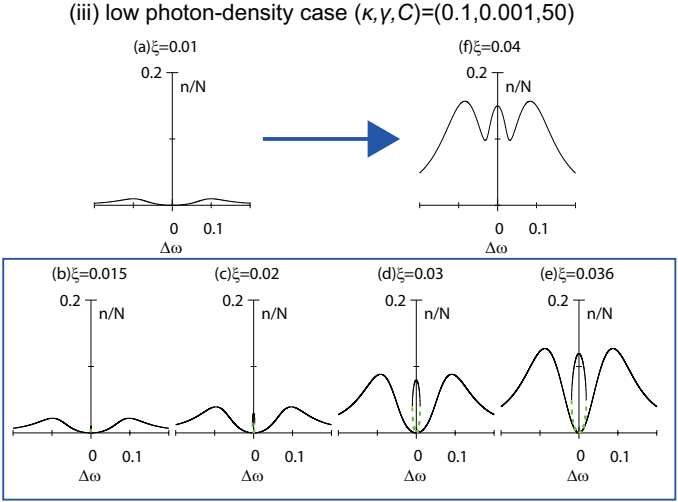}
\caption{
Dependence of the photon number density $n/N$ on the detuning frequency $\Delta \omega$ in the low photon-density case, ($\kappa, \gamma, C) = (0.1, 0.001, 50$), where the clear two peaks at finite $\Delta \omega$ remain even at $\xi=0.04$.
The transition of the structures from (a) $\xi=0.01$ to (f) $\xi=0.04$ is shown in the box: (b) $\xi =0.015$, (c) $\xi =0.02$, (d) $\xi =0.03$, and (e) $\xi =0.036$.
The stable and unstable MF steady-state solutions are denoted by solid lines (black) and dotted lines (green), respectively.
}
\label{MF-detuning3}
\end{figure}

\begin{figure}[t]
\includegraphics[clip,width=0.5\textwidth]{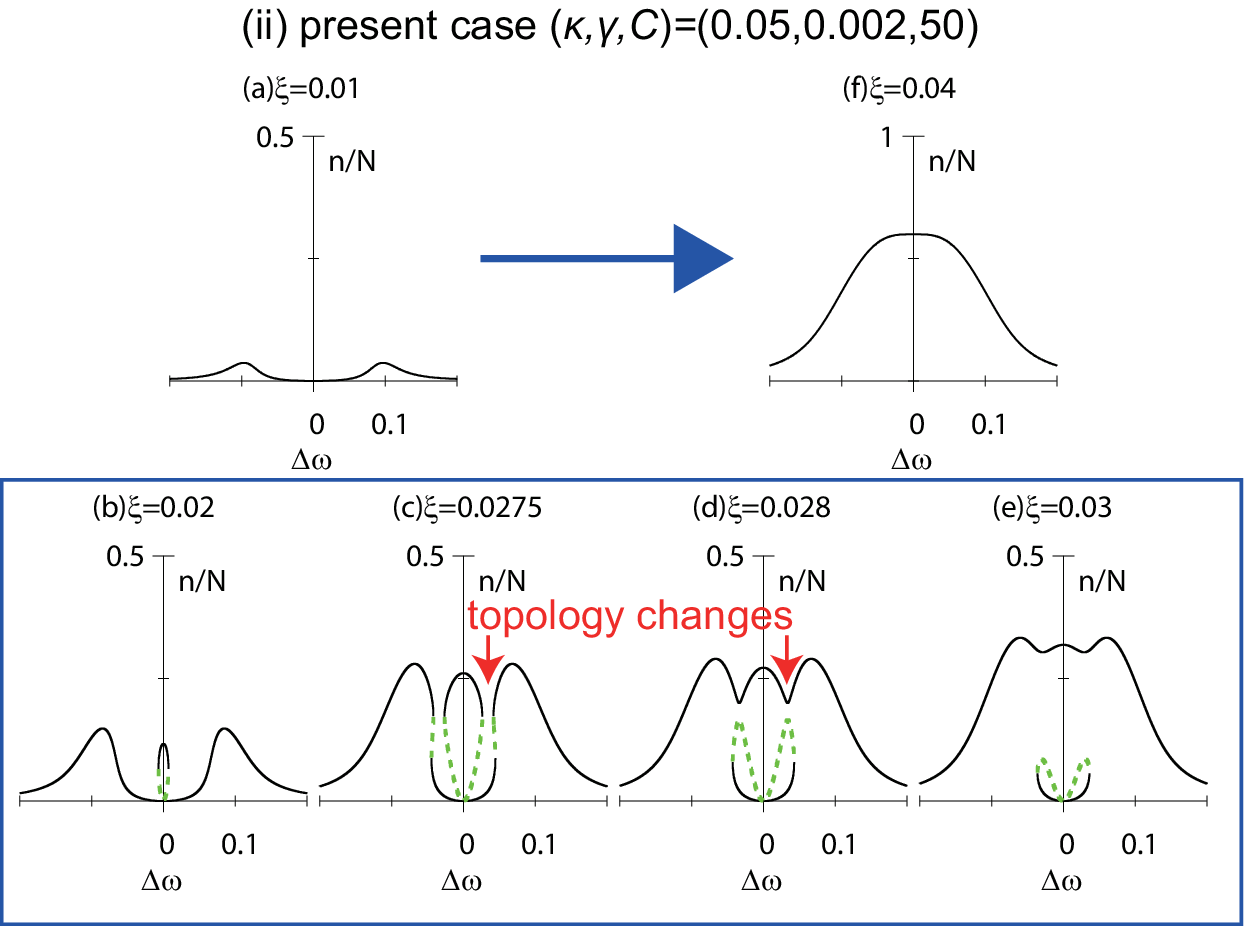}
\caption{
Dependence of the photon number density $n/N$ on the detuning frequency $\Delta \omega$ in the present case, ($\kappa, \gamma, C) = (0.05, 0.002, 50$).
The transition of the structures from (a) $\xi=0.01$ to (f) $\xi=0.04$ is shown in the box: (b) $\xi =0.02$, (c) $\xi =0.0275$, (d) $\xi =0.028$, and (e) $\xi =0.03$.
Between (c) and (d), the topology of $n(\Delta\omega)$ changes.
The stable and unstable MF steady-state solutions are denoted by solid lines (black) and dotted lines (green), respectively.
}
\label{MF-detuning}
\end{figure}

In this section we discuss the dependencies of the steady-state solutions on the detuning frequency $\Delta \omega$ [see Eq.~(\ref{detune_eq})],
and show characteristic properties of the present optical bistability within the low photon-density regime.

At nonzero $\Delta \omega$, the relation between $\xi_{\rm s}$ and $\alpha_{\rm s}$ in the MF analysis is given by
\begin{align}
\left(\frac{\xi_{\rm s}}{\alpha_{\rm s}}\right)^2=&\left(1+\frac{2C}{1+\alpha_{\rm s}^2+(\frac{\Delta\omega}{\gamma})^2}\right)^2\nonumber\\
&+\left(\frac{\Delta \omega}{\kappa}\right)^2 \left(1-\frac{\kappa}{\gamma}\frac{2C}{1+\alpha_{\rm s}^2+(\frac{\Delta\omega}{\gamma})^2}\right)^2,
\end{align}
which is reduced to Eq.~(\ref{OB_scale}) in the resonant case, i.e., $\Delta \omega=0$.
The photon number density in the MF treatment is given by
\begin{equation}
\frac{n(\Delta\omega)}{N}\equiv \frac{1}{N}{\rm Tr} (a^\dagger a \rho_{\rm ss})=\frac{\gamma}{4\kappa C}\alpha_{\rm s}^2.
\end{equation}
In contrast to the resonant case, the dependencies between $\alpha_{\rm s}$ and $\xi_{\rm s}$ at finite $\Delta \omega$ are not only determined by $C$ but by all the parameters: $\kappa, \gamma$, and $C$.
Thus, the structures of the MF steady-state solutions in $\Delta \omega$-$n$ plane depend on $\kappa$ and $\gamma$ even when $C$ is the same.

We study how the structures of $n(\Delta \omega)$ in the MF analysis depend on the dissipation rates, $\kappa$ and $\gamma$, which controls the photon number density.
The transmission spectrum, i.e., $n(\Delta \omega)$, shows a single peak in the high photon-density regime,
while it shows a double peak in the low photon-density regime.
Thus, it is expected that the MF steady-state solutions extend to $\Delta \omega$ direction in a different manner depending on whether the state is in the high photon-density regime or the low photon-density regime.

We consider three cases: (i) standard case ($\kappa,\gamma,C$)$=$($0.01,0.01,50$), (ii) present case ($\kappa,\gamma,C$)$=$($0.05,0.002,50$), and (iii) low photon-density case ($\kappa,\gamma,C$)$=$($0.1,0.001,50$).
In case (i), the high photon-density state of the optical bistability is in the high photon-density regime, i.e., $n/N>1$ [see Fig.~\ref{MF-detuning2}(f)].
On the other hand, in cases (ii) and (iii) [see Fig.~\ref{MF-detuning}(f) and Fig.~\ref{MF-detuning3}(f)], they are in the low photon-density regime, i.e., $n/N<1$.
The structures of $n(\Delta\omega)$ are qualitatively different in the three cases (i), (ii), and (iii), as is shown below.
It is noted that we set $C$ and $\kappa\gamma$ to be the same and therefore for all the three cases, the bistable MF solutions appear at $\xi =\xi_{\rm l}$ and disappear at $\xi_{\rm u}$ in the resonant case, i.e., $\Delta \omega =0$ [see Eqs.~(\ref{scale})-(\ref{cooperativity}), and Fig.~\ref{OB}].

The transmission spectrum in case (i) [the standard case, $(\kappa, \gamma, C)=(0.01, 0.01, 50)$] is depicted in Fig.~\ref{MF-detuning2}.
From the low photon-density regime [Fig.~\ref{MF-detuning2}(a)] to the high photon-density regime [Fig.~\ref{MF-detuning2}(f)],
the double peak changes to the single peak, which was observed in a cavity QED experiment~\cite{gripp1996anharmonicity}.
Between them, first the two branches of the double peak develop with $\xi$ and then the branches merge at $\Delta \omega =0$ at $\xi_{\rm l}$ [Fig.~\ref{MF-detuning2}(b)].
At the merging point a loop appears, and consequently the topological structure of $n(\Delta \omega)$ changes.
As $\xi$ further increases, the loop shrinks and then disappears at $\xi_{\rm u}$ [Fig.~\ref{MF-detuning2}(b)-(f)].

The transmission spectrum in case (iii) [the low photon-density case, $(\kappa, \gamma, C)=(0.1, 0.001, 50)$] is depicted in Fig.~\ref{MF-detuning3}.
The double peak at $\xi=0.01$ [Fig.~\ref{MF-detuning3}(a)] remains visible even at $\xi=0.04$ [Fig.~\ref{MF-detuning3}(f)].
Between them, first a narrow loop appears along $n/N$-axis at $\xi_{\rm l}$ [Fig.~\ref{MF-detuning3}(b)].
Then the width of the loop increases with $\xi$ [Fig.~\ref{MF-detuning3}(b)-(e)], and at last the unstable MF solutions denoted by dotted lines (green) merge with the double peak at $\xi_{\rm u}$ [Fig.~\ref{MF-detuning3}(e)].
It is noted that the way of emerging and merging the loop gives a different topological structure from case (i) (see Fig.~\ref{MF-detuning2}).  

The transmission spectrum in case (ii) [the present case, $(\kappa, \gamma, C)=(0.05, 0.002, 50)$] is depicted in Fig.~\ref{MF-detuning}.
The double peak at $\xi=0.01$ [Fig.~\ref{MF-detuning}(a)] disappears at $\xi=0.04$ [Fig.~\ref{MF-detuning}(f)] even though the state is in the low photon-density regime.
The transition between them shows again another topological structure.
Namely, first a narrow loop appears along $n/N$-axis at $\xi=\xi_{\rm l}$ [Fig.~\ref{MF-detuning}(b)] similar to case (iii).
As $\xi$ increases, the loop merges with the double peak between $\xi=0.0275$ [Fig.~\ref{MF-detuning}(c)] and $\xi=0.028$ [Fig.~\ref{MF-detuning}(d)] in a different manner as in Fig.~\ref{MF-detuning3} and the topology of $n(\Delta \omega)$ changes at this point.
It is noted that the point is rather close to the crossing point $\xi_{\rm c}\simeq 0.029$ in Fig.~\ref{OB} although the relation is so far unclear.
After that, the topology of $n(\Delta \omega)$ is similar to that of the standard case, Fig.~\ref{MF-detuning2}, and the loop shrinks with the increase of $\xi$ and disappears at $\xi_{\rm u}$ [Fig.~\ref{MF-detuning}(d)-(f)].

In experiment, the differences among the three types of topological structures will appear in the way how $n(\Delta \omega)$ changes from $\Delta\omega=0$ to a non-zero value of $\Delta\omega$ in the bistable regime, i.e., $\xi_{\rm l} \leq \xi \leq \xi_{\rm u}$.
First, suppose that the system is in the high photon-density state of the optical bistability at $\Delta\omega=0$.
Here, $n(\Delta\omega)$ continuously decreases with $\Delta\omega$ in case (i), while it shows a discontinuous jump to the low photon-density state at a certain value of $\Delta\omega$ in cases (ii) and (iii).
On the other hand, when the system is in the low photon-density state at $\Delta\omega=0$,
$n(\Delta \omega)$ shows a discontinuous jump to the high photon-density state in cases (i) and (ii), while it follows a continuous curve in case (iii).
Whether the state at $\Delta\omega=0$ is in the low photon-density state or the high photon-density state depends on the value of $\xi$ (see Fig.~\ref{OB}).

\begin{figure}[t]
\includegraphics[clip,width=0.5\textwidth]{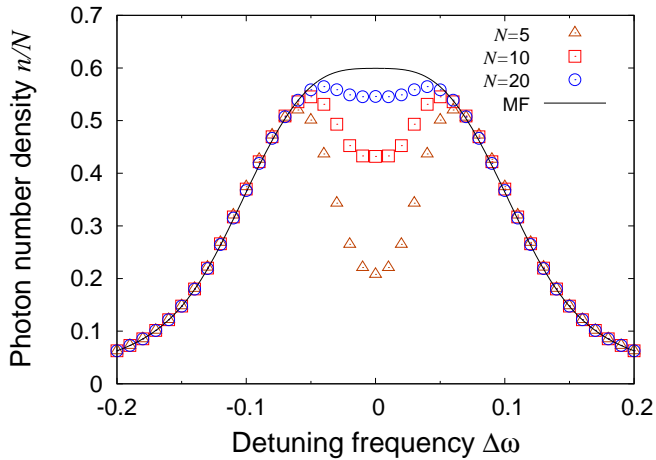}
\caption{Size dependence of the photon number density on the detuned driving frequency $\Delta \omega$ in systems with $N=5$ (triangles), $10$ (squares), $20$ (circles), and $\infty$ (MF) at $\xi=0.04$.}
\label{detune_size}
\end{figure}

Finally, we study the detuning effects in finite systems with parameters.
In Fig.~\ref{detune_size}, we show the size dependence of $n/N$ on $\Delta\omega$ for $\xi=0.04$.
In finite systems, the double peak structure is more clearly visible.
Although only a single peak appears in the MF solution [see Fig.~\ref{MF-detuning}(f)], we observe a clear double peak structure for $N=5$.
The steady-state solution approaches the MF result as $N$ increases.

\section{Summary and Discussion}\label{summary}
In the present paper, we have studied the properties of the optical bistability for systems up to $25$ atoms in the low photon-density regime, where the photon number density $n/N$ is less than one.
Although the static properties in the thermodynamic limit can be obtained by the MF treatment, the phenomena in finite systems are interesting for micro-size quantum manipulations.
We studied the phenomena in systems with finite number of atoms by a numerically exact method.
We characterized the phenomena in terms of the eigenmodes and the eigenvalues of the time evolution operator of the QME, ${\cal L}$ [Eq.~(\ref{QME_R})].

We developed an efficient numerical scheme to treat the QME for hybridized systems of photons and a large number of two-level atoms.
This scheme consists of the parallelization in photon space and the reduction of the Hilbert space of atoms.
We confirmed the good efficiency of the parallelization (see Appendix~\ref{APP2}).
Note that the limitation of system size up to $N=25$ in the present study is not due to the memory to store the density matrix, but due to the computational time to estimate the steady-state density matrix in the bistable regime.
The significantly small eigenvalue of ${\cal L}$ in the bistable regime makes the convergence of the Bi-CG method worse, which leads to the increase of the computational time.

We investigated the size dependence of the photon number density as a function of the amplitude of the driving field, and there we found that the steady state values quickly approach the MF values outside the bistable regime (Fig.~\ref{OB}).
Inside the bistable regime, we found a crossing point for different system sizes.
Around this point, we analyzed the density matrix of the steady state, which is the eigenmode of ${\cal L}$ with zero eigenvalue.
We found that the double peak structure appears around the crossing point and the size dependence of the double peak structure changes at this point (Figs.~\ref{photon_distribution} and \ref{distribution}).

We also studied dynamical properties.
We characterized the timescale for relaxation by the gap of the eigenvalues of ${\cal L}$ [see Eq.~(\ref{relaxation})], and found the exponential growth of the relaxation time in the bistable regime as $N$ increases (Fig.~\ref{gap}).
The signature of the long timescale appears in the scaling form of the hysteresis loop (Fig.~\ref{hysteresis}).

In the present study, we concentrated on the low photon-density regime.
The qualitative difference from the standard optical bistability appears in the transmission spectrum as a function of the detuning frequency $\Delta \omega$, i.e., $n(\Delta \omega)$.
We found three different types of the transmission spectrum, $n(\Delta \omega)$, depending on the dissipation rates, $\kappa$ and $\gamma$ (Figs.~\ref{MF-detuning2}-\ref{MF-detuning}).

It would be an interesting problem in the future to study the crossing point $\xi_{\rm c}$ in the limit of $N$ to infinity.
We showed that the free energy landscape estimated by the MF solution does not allow to obtain the crossing point (see Sec.~\ref{sec:effective}).
Moreover, the effect of short-range interaction between spins, the dipole-dipole interaction, on the steady-state density matrix $\rho_{\rm ss}$ and the relaxation time $\tau$ is an important issue to be investigated in near future.

\begin{acknowledgments}
  This research was supported by MEXT as ``Exploratory Challenge on Post-K computer'' (Challenge of Basic Science -- Exploring Extremes through Multi-Physics and Multi-Scale Simulations).
  The numerical calculations in the present work have been done mainly on the K computer at RIKEN R-CCS and the supercomputer system at Institute for Solid State Physics, University of Tokyo.
We thank Nobuyasu Ito for his suggestion on the optimization of core allocation on the K computer.
\end{acknowledgments}

\appendix

\section{Efficiency of the parallelization}~\label{APP2}
\begin{figure}[t]
\includegraphics[clip,width=0.5\textwidth]{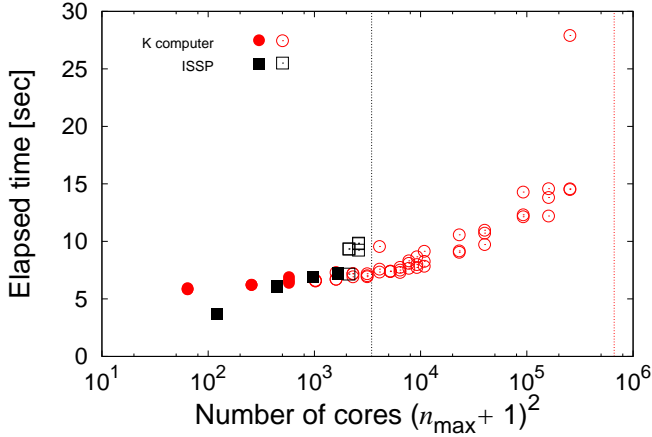}
\caption{
Elapsed time for the $10000$ multiplications of ${L}$ and $L^{\dagger}$ [Eq.~(\ref{matrix_rep})].
Data taken on the ISSP supercomputer system are plotted by the black squares and those on the K computer are plotted by red circles.
Filled symbols denote the case where the simulation fits in a single unit of the computers, while empty symbols denote the case where the data exchanges between the units are necessary.
The black and red vertical dotted lines denote the upper limit of the number of cores, $3,456$ for the ISSP supercomputer system and $663,552$ for the K computer, respectively.
For each data point, the elapsed time is measured three times.
}
\label{efficiency}
\end{figure}

\begin{figure}[t]
\includegraphics[clip,width=0.5\textwidth]{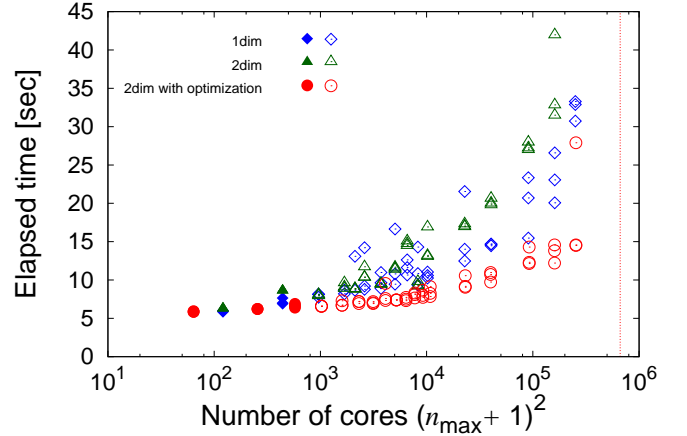}
\caption{
Elapsed time for the $10,000$ multiplications of ${L}$ and $L^{\dagger}$ [Eq.~(\ref{matrix_rep})].
Data taken on the K computer with the one-dimensional core allocation are plotted by blue diamonds.
Those with two-dimensional core allocation are plotted by green triangles.
The red ones are the results of the case of optimized core allocation.
Filled symbols denote the case where the simulation fits in a single unit of the computers, while empty symbols denote the case where the data exchanges between the units are necessary.
The red vertical dotted line denotes the upper limit of the number of cores, $663,552$ for the K computer.
For each data point, the elapsed time is measured three times.
}
\label{efficiency2}
\end{figure}

We study the efficiency of the parallelization in terms of the photon space.
The core labeled by the pair of integers $(n_1, n_2)$ stores elements of $\bra{n_1} \rho \ket{n_2}$, where $\{ \ket{n} \}$ are photon number states and the integer $n$ runs from $0$ to the cutoff $n_{\rm max}$.
Thus, the total number of cores is
\begin{equation}
n_{\rm core}=(n_{\rm max}+1)^2.
\end{equation}
The main part of the numerical calculation is the Bi-CG method, consisting of the multiplication of $L$ and $L^{\dagger}$ on $\vec{\rho}$.
The calculation of $(n_1, n_2)$-elements of $L \vec{\rho}$ and $L^{\dagger} \vec{\rho}$ requires only six elements of $\rho$ as depicted in Fig.~\ref{cores}.
Due to the local nature of the calculation independent of $n_{\rm max}$, good efficiency should be achieved.

We consider the weak scaling to evaluate the efficiency of the parallelization.
Namely, we fix the number of atoms to be $N=10$ while increasing $n_{\rm max}$, and calculate the elapsed time for $10000$ multiplications of ${L}$ and $L^{\dagger}$ to $\vec{\rho}$.
We plot the result of benchmark test on the supercomputer system (SGI ICE XA/UV) at ISSP, University of Tokyo and the K computer at RIKEN R-CCS in Fig.~\ref{efficiency}.
In this figure, we use the filled symbols when $n_{\rm core}$ is less than the number of cores in a single unit, i.e., $n_{\rm ISSP} = 1728$ (72 nodes) and $n_{\rm K} = 768$ (96 nodes) for the ISSP supercomputer system and the K computer, respectively,
and we use open symbols for the cases with the larger $n_{\rm core}$.
We run the same jobs for each $n_{\rm max}$ three times and plot them.
We could simulate up to $n_{\rm core}=3456$ $(n_{\rm max}\simeq 57)$ and $n_{\rm core}=663,552$ $(n_{\rm max}\simeq 800)$ on the ISSP system and the K computer, respectively.  In Fig.~\ref{efficiency}, we indicate the maximum number of cores for each machine by the vertical lines.

In Fig.~\ref{efficiency}, the elapsed time in both machines exhibit a plateau (filled symbols), i.e., almost ideal weak scaling as long as $n_{\rm core}$ is smaller than or equal to the number of cores in a single unit, $n_{\rm ISSP}$ or $n_{\rm K}$. 
However, once $n_{\rm max}$ exceeds $n_{\rm ISSP}$, the elapsed time shows a sudden growth in the case of the ISSP system (open squares).
The increase of the elapsed time may be due to the data exchange between different units.
The elapsed times of the K computer, on the other hand, stays flat even at $n_{\rm core}=10^4$ (open circles), even though $n_{\rm core}$ is significantly larger than $n_{\rm K}$.
This indicates the higher performance of communication between different units of the K computer.
The increase of the elapsed time for $n_{\rm core}>10^4$ may be improved if we use the MPI/OpenMP hybrid parallelization instead of the present flat MPI scheme, which is an issue to be examined in the future.

We also find the strong dependencies of the elapsed time on the way cores labeled by $(n_1, n_2)$ are allocated on the K machine, as shown in Fig.~\ref{efficiency2}.
We find that the data start to fluctuate considerably when $n_{\rm core}$ exceeds the number of cores in a single unit, $n_{\rm K}$.
If we allocate cores in the so-called one-dimensional way, the average elapsed times are much larger (blue diamonds).
Even if we allocate cores in the so-called two-dimensional way, the situation is not improved (green triangles).
If we allocate cores so that they are closer when the indices $n_1$ and $n_2$ are close, the performance is much improved (red circles, which are also plotted in Fig.~\ref{efficiency}).

\bibliography{bistability.bib}

\end{document}